\documentclass[conference,a4paper]{APSIPA2025}
\usepackage{amsmath,amssymb,amsfonts,bm,cite,mleftright,xargs,siunitx}
\usepackage{graphicx}
\usepackage{multirow}
\usepackage{threeparttable}

\usepackage{booktabs} 
\usepackage{subfigure} 

\usepackage{makecell}
\usepackage{hyperref}

\usepackage{geometry}
\usepackage{fancyhdr}

\fancypagestyle{firststyle}{
  \fancyhf{}
}

\begin{document}

\title{Efficient Transformer-Based 
Piano Transcription With Sparse Attention Mechanisms}


\fussy

\author{
\authorblockN{
Weixing Wei\authorrefmark{1} and
Kazuyoshi Yoshii\authorrefmark{1}
}

\authorblockA{
\authorrefmark{1}
Kyoto University, Japan \\
E-mail: wei.weixing.23w@st.kyoto-u.ac.jp, yoshii.kazuyoshi.3r@kyoto-u.ac.jp}


}

\maketitle
\thispagestyle{firststyle}
\pagestyle{empty}





\begin{abstract}

This paper investigates automatic piano transcription
 based on computationally-efficient
 yet high-performant variants of the Transformer
 that can capture longer-term dependency
 over the whole musical piece.
Recently, transformer-based sequence-to-sequence models
 have demonstrated excellent performance 
 in piano transcription. 
These models, however, 
 fail to deal with the whole piece at once 
 due to the quadratic complexity of the self-attention mechanism,
 and music signals are thus typically processed 
 in a sliding-window manner in practice.
To overcome this limitation, 
 we propose an efficient architecture
 with sparse attention mechanisms. 
Specifically, we introduce sliding-window self-attention mechanisms
 for both the encoder and decoder,
 and a hybrid global-local cross-attention mechanism
 that attends to various spans according to the MIDI token types.
We also use a hierarchical pooling strategy
 between the encoder and decoder
 to further reduce computational load.
Our experiments on the MAESTRO dataset showed that
 the proposed model achieved a significant reduction 
 in computational cost and memory usage, 
 accelerating inference speed, 
 while maintaining transcription performance comparable 
 to the full-attention baseline.
This allows for training with longer audio contexts
 on the same hardware, demonstrating the viability
 of sparse attention for building efficient 
 and high-performance piano transcription systems.
The code is available at \url{https://github.com/WX-Wei/efficient-seq2seq-piano-trans}.
\end{abstract}


\section{Introduction}

Automatic music transcription (AMT) is one of the most fundamental tasks that aims to convert raw audio into symbolic musical notation \cite{AMT-overview/benetos2018}, such as MIDI or music score. 
Automatic piano transcription, a subtask of AMT, has received significant attention due to the wide range and polyphonic nature of frequency components.
Recent years have seen a paradigm shift towards deep learning-based approaches for AMT. 
Convolutional recurrent neural networks (CRNNs) \cite{Onsets_Frames,high_resolution,hppnet-Wei2022} have demonstrated high performance
 by effectively capturing local time-frequency patterns and their temporal dependencies. 
These models typically operate at a frame level,
 predicting which notes are active at each short time step.

More recently, sequence-to-sequence models
 based on the Transformer \cite{attention-is-all-you-need/nips/VaswaniSPUJGKP17} architecture have been applied successfully to this task \cite{transformer,MT3/iclr/GardnerSMHE22,exploring-transformer/icassp/OuGBHW22},
 leveraging its ability to transform music transcription tasks
 into a sequence-to-sequence problem.
This approach directly translating an input audio representation 
(e.g., short-time Fourier transform or mel spectrogram)
 into a sequence of symbolic music events.
The Transformer has proven to be highly effective for this task,
 establishing comparable performance to state-of-the-art frame-level automatic music transcription models.

\begin{figure}[t]
    \begin{center}
    \includegraphics[width=90mm]{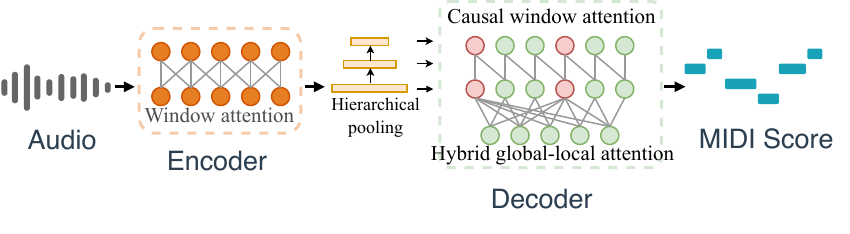}
    \end{center}
    \vspace{-3mm}
    \caption{Diagram of the efficient transformer for piano transcription. 
    Local sliding window attention, hybrid global-local attention mechanisms, and hierarchical pooling are proposed to reduce the computational cost.} %
    \vspace*{-20pt}
    \label{fig:efficient-transformer-workflow}
\end{figure}

Despite its success, the vanilla Transformer has a significant drawback. 
The computational and memory complexity of its self-attention mechanism
 is quadratic with respect to the input sequence length.
This becomes one of the major bottlenecks in automatic music transcription,
 where both the input audio and output MIDI sequences can be very long,
 especially when high temporal resolution is desired.

A key finding through 
 our preliminary analysis of a vanilla Transformer-based 
 audio-to-MIDI piano transcription
 is that the attention weights are remarkably sparse. 
This sparsity is observed 
 in the self-attention mechanism of the encoder and decoder
 and the encoder-decoder cross-attention mechanism.
This phenomenon suggests that the relationship 
 between the input audio and the output MIDI representation 
 has strong local correspondence.
A specific MIDI event (e.g., note onset) is most strongly
 related to the audio features in a narrow time window around it.
The global, long-range dependencies 
 across the time axis appear to be weaker.
This insight motivates our work. 
We hypothesize that the dense, full attention mechanism 
 is not necessary for piano transcription
 and can be replaced with a more efficient sparse one 
 without a significant loss in accuracy. 

In this work, we propose a Transformer-based piano transcription model
 with sliding window self-attention
and a novel hybrid global-local cross-attention scheme
 to leverage the local nature of acoustic-to-symbolic correspondence. 
We also introduce a hierarchical pooling mechanism
 between the encoder and the decoder to balance time resolution and computational efficiency. 
The hybrid attention selectively applies global attention only to time-positioning tokens
 while using local attention for note and velocity tokens. Fig. \ref{fig:efficient-transformer-workflow} shows the framework of the model.
Our model achieves comparable transcription performance to full-attention Transformers, 
 while significantly reducing inference time and computational cost.



\section{Related Work}

This section reviews recent attempts
 to improve the performance and efficiency of piano transcription.




\subsection{Improving Performance of Piano Transcription}

Traditional machine learning approaches for AMT formulate the task as a frame-level classification problem. 
A model processes an input spectrogram and outputs a multi-label classification for each time frame,
 indicating which of the 128 MIDI notes are active. 
CRNN-based architectures have become a popular and effective choice for piano transcription.
These models,
 pioneered by the "Onsets and Frames" \cite{Onsets_Frames} achieved significant progress.
The High-Resolution Piano Transcription (HPT) model \cite{high_resolution} performed regression on note and pedal event detection, demonstrating high-precision time resolution transcription performance. 
While successful, these frame-level transcription models typically formulate the task as a multi-label classification problem at each time frame,
 assuming that frames are conditionally independent.
This assumption neglects the temporal dependencies between frames,
 which are crucial for accurate music transcription. 
Although recurrent neural networks (RNNs) have been employed to capture temporal context,
 they have shown limited effectiveness
 in modeling the temporal structure required to accurately associate note onsets and offsets.
As a result, these frame-level CRNN based models often struggle to detect note offset events.

To capture time-frequency correlation,
 several attempts have been made to introduce
 time and frequency-domain attention mechanisms 
 into frame-level models.
For instance, the hFT-Transformer \cite{hft-transformer-toyama2023} 
 introduced a combination 
 of frequency- and time-wise attention mechanisms
 for better structural modeling of musical information. 
A harmonic frequency and time attentions were
 proposed to capture musical frequency structure and temporal dependencies
 \cite{WangLBJ24-HarmonicawareFrequencyTime}.
Incorporating a local relative time attention mechanism
 \cite{WangLCX23-local-rel-attt-piano} into the CNN encoder, 
 or using Transformer decoder as the adapter 
 between the CNN and RNN
 \cite{DBLP:conf/apsipa/MiKT24-improving-high-res-piano-transcription} 
 also show transcription performance improvement.


Besides the frame-level models, 
 sequence-to-sequence models have emerged as a powerful alternative, 
 and later refined with Transformers \cite{transformer, MT3/iclr/GardnerSMHE22}.
The sequence-to-sequence approach
 directly generate a symbolic representation,
 such as a sequence of MIDI-like tokens. 
This sequence-to-sequence approach elegantly integrates onset, offset, and velocity prediction into a single generative process,
 simplifying piano transcription into a simple standard task. 
Our work builds upon this sequence-to-sequence paradigm.

Other methods have incorporated semi-Markov conditional random fields (semi-CRFs) \cite{skip_frame, yan2024scoring}  on top of neural network outputs to enforce musically meaningful constraints on the predicted note sequence. 
Although achieving state-of-the-art piano transcription performance, these works often led to the cost of increased model complexity.

\subsection{Improving Efficiency of Piano Transcription}

In parallel, another line of research has aimed to improve the computational efficiency of piano transcription models. 
This includes the development of lightweight models like HPPNet\cite{hppnet-Wei2022} and lightweight instrument-agnositc model proposed in \cite{BittnerBRME-ICASSP22-a-lightweight-instrument},  streaming piano transcription model \cite{Wei24-streaming-piano} designed for low-latency performance. 
Other efforts have focused on creating real-time transcription systems \cite{DBLP:conf/eusipco/Fernandez23-onsets-and-velocities-affordable-real-time, ismir/jeong2020real-time-piano-transcription}, by simplifying model architectures and optimizing inference pipelines. 
Our work aligns with this goal of efficiency,
  but focuses specifically on optimizing the Transformer architecture itself.



The quadratic complexity of the vanilla Transformer's attention mechanism has been a major research topic in the broader machine learning community. 
This has led to the development of numerous "Efficient Transformer" variants that employ sparse attention patterns. 
Models like Longformer \cite{beltagy2020longformer} and BigBird \cite{zaheer2020big-bird-nips2020} use a combination of local windowed attention and global attention mechanisms on selected tokens to approximate full attention. 
Other methods, such as FlashAttention\cite{dao2022flashattention}, optimize the computation of attention at the hardware level. 
Our work draws inspiration from these general techniques,
  adapting and specializing the concept of sparse attention mechanism for the specific domain of piano music transcription.

\section{Model Architecture}

\begin{figure}[t]
    \begin{center}
    \includegraphics[width=80mm]{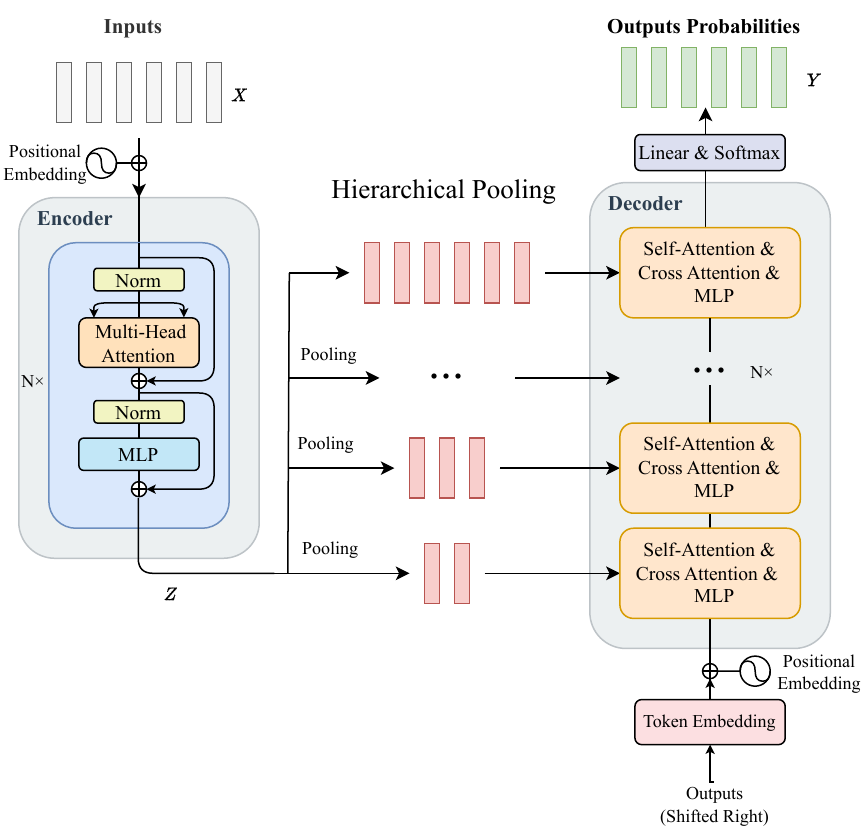}
    \end{center}
    \vspace{-3mm}
    \caption{Overview of the proposed efficient Transformer architecture. 
    Both the encoder and decoder employ local 
    sliding-window self-attention mechanisms 
    to reduce computational complexity. 
    The encoder output is pooled with different kernel sizes 
    and fed into different layers of the decoder. 
    The encoder-decoder cross-attention 
    adopts a hybrid global-local mechanism.}
    \vspace*{-3pt}
    \label{fig:efficient-transformer}
\end{figure}

Our baseline model is a sequence-to-sequence Transformer architecture, similar to the Text-to-Text Transfer Transformer (T5) \cite{T5-RaffelSRLNMZLL20}. 
As illustrated in Fig.~\ref{fig:efficient-transformer}, the model consists of an encoder and a decoder. 
The encoder maps a sequence of input audio features $X = (x_1, ..., x_T)$ to a sequence of latent representations $Z = (z_1, ..., z_T)$. 
The decoder then autoregressively generates an output sequence of symbolic music tokens $Y = (y_1, ..., y_N)$ based on $Z$ and the previously generated tokens.
We follow \cite{transformer}, using the same T5 architecture for our transcription model.
Each encoder and decoder block uses a pre-layer normalization (Pre-LN) structure and consists of two kinds of sub-layers: 
 the multi-head attention (MHA) mechanism and the position-wise feed-forward network (FFN). 
Residual connections are applied around each sub-layer. Dropouts are applied immediately after each MHA and FFN layers.
The activation function used is ReLU.


\subsection{MIDI-Like Tokenization}

For audio-to-MIDI (sequence-to-sequence) piano transcription, 
 we tokenize the ground-truth MIDI performance 
 into a linear sequence of events.
Events are ordered strictly by their onset times.
For simultaneous events (e.g., simultaneous note onsets),
 they are further sorted by pitch in ascending order.
Our vocabulary consists of the following token types:


\begin{itemize}
    \item \textbf{Time}: 600 values that represent absolute time locations quantized into 600 discrete bins. With a hop length of 20 ms, this allows for representing time up to 30 seconds.
    \item \textbf{Note On/Off}: 128 pitch classes each that represent the onset and offset of one of the 128 MIDI notes
    \item \textbf{Velocity}: 128 velocity levels.
    \item \textbf{Special Tokens}:  Three values that include BOS (Begin of Sequence), EOS (End of Sequence), and PAD (Padding).
\end{itemize}

\begin{figure}[tb]
\centering
    \subfigure[Sliding window attention]{
        \includegraphics[width=2.2cm]{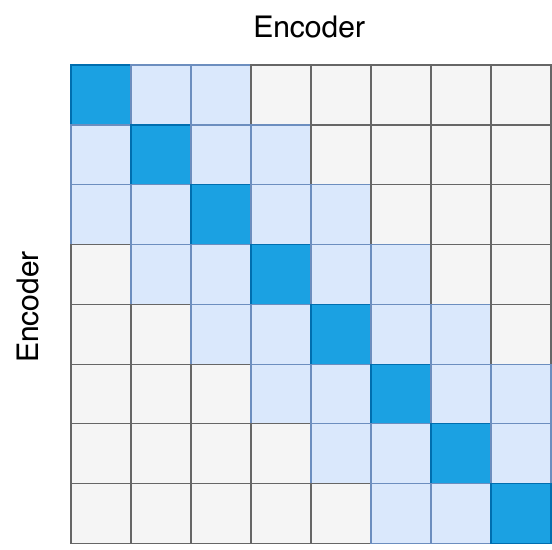}
        \label{fig:sparse-attention-a}
    }
    \hspace{0.1in} 
    \subfigure[Hybrid global-local attention]{
        \includegraphics[width=2.2cm]{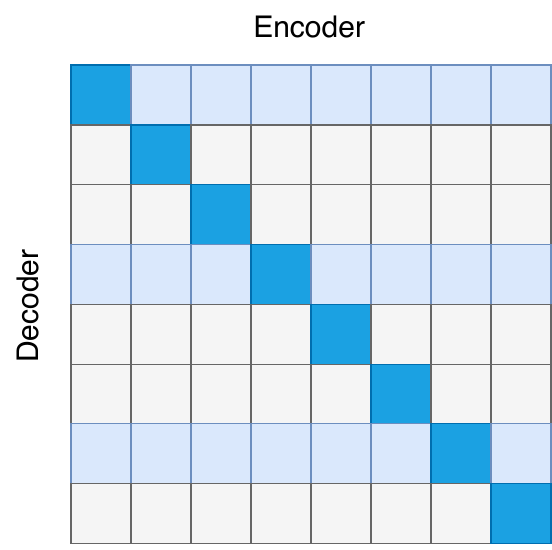}
        \label{fig:sparse-attention-b}
    }
    \hspace{0.1in} 
    \subfigure[Causal sliding window attention]{
        \includegraphics[width=2.2cm]{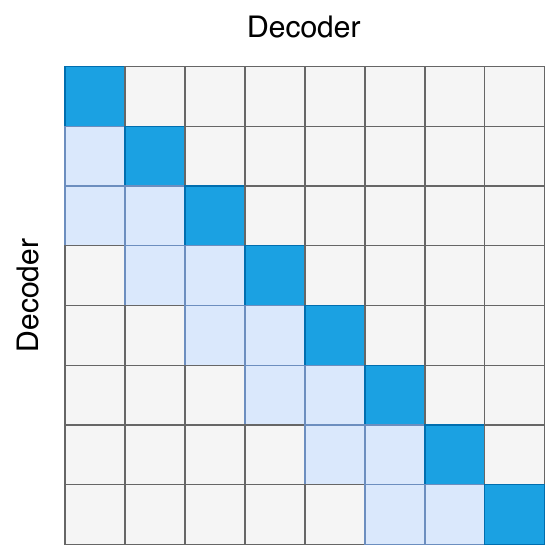}
        \label{fig:sparse-attention-c}
    }

\caption{An illustration showing a banded sliding window self-attention mask for the encoder on the left (a),
 a hybrid global-local attention mask for encoder-decoder attention on the middle (b),
 and a causal sliding window self-attention for the decoder on the right (c).}
\end{figure}

\subsection{Sliding Window Attention}

The core computation in the Transformer is the scaled dot-product attention \cite{attention-is-all-you-need/nips/VaswaniSPUJGKP17}.
It computes attention scores by taking the dot product between query and key vectors, 
 then scaling the result by a scale factor to stabilize gradients during training. 
Mathematically, the attention weights are computed as:

\begin{equation}
\text{Attention}(Q, K, V) = \text{softmax}\left(\frac{QK^T}{\sqrt{d_k}}\right)V,
\end{equation}
where $Q$, $K$, and $V$ are the query, key, and value matrices
 and $d_k$ is the dimension of the key vectors. 

For a sequence of length $N$, the complexity of this operation is $\mathcal{O}(N^2 \cdot d_k)$, 
 which is computationally expensive for long sequences. 
 Since $d_k$ is a hyperparameter determined by the model architecture and remains constant regardless of input length, the complexity simplifies to $\mathcal{O}(N^2)$ in practice.

Based on our observation of attention sparsity in the piano transcription transformer, 
 we replace the full self-attention mechanism in both the encoder and decoder with a sliding window self-attention mechanism. 
In this setting, each token attends only to its immediate neighbors within a fixed window size $w$. 
This can be implemented by applying a mask to the attention score matrix $QK^T$ before the softmax operation as demonstrated in Eq. (\ref{eq:sliding-window-mask}). 
The attention field is restricted to a band matrix, 
 where only the elements $A_{i,j}$ with $|i-j| \le w/2$ are computed. In the decoder, each token is generated sequentially based on previous tokens, requiring strict causality. 
The mask matrix is modified to Eq. (\ref{eq:causal-sliding-window-mask}), the attention calculation is modified to Eq. (\ref{eq:mask-attention}).
This allows the model to compute attention weights only within the sliding window.
This modification from full attention to sliding window attention
 reduces the computational complexity from $\mathcal{O}(N^2)$ to $\mathcal{O}(N \cdot w)$, where $w \ll N$. 
Fig.~\ref{fig:sparse-attention-a} and \ref{fig:sparse-attention-c} illustrates this concept. 
From a theoretical perspective, the masked positions (where \( M_{ij} = -\infty \)) can be entirely skipped during computation. 
In practice, to fully exploit hardware-level parallelism, we represent masked positions using a mask matrix 
 and employ Flash Attention\cite{dao2022flashattention}, 
 which accelerates attention through an efficient block-wise computation strategy.


\begin{equation}
    M_{i,j} = \begin{cases}
    0 & \text{if } |i - j| \leq w/2 \\
    -\infty & \text{otherwise}
    \end{cases},
    \label{eq:sliding-window-mask}
\end{equation}
\begin{equation}
    M_{i,j} = \begin{cases}
    0 & \text{if }  0 \leq (i-j) \leq w \\
    -\infty & \text{otherwise}
    \end{cases},
    \label{eq:causal-sliding-window-mask}
\end{equation}
\begin{equation}
    \text{LocalAttention}(Q, K, V) = \text{softmax}\left(\frac{QK^T + M}{\sqrt{d_k}}\right)V,
    \label{eq:mask-attention}
\end{equation}
where $M$ is the mask matrix, $i$ is the query sequence index, $j$ is the key sequence index.


\subsection{Hybrid Global-Local Attention}

While sliding window attention is effective for self-attention layers, 
 the encoder-decoder cross-attention presents a different challenge. 
In cross-attention, 
 the decoder's queries attend to the encoder's keys and values. 
We hypothesize that not all types of tokens
 in the decoder need a full attention view of the entire encoded audio sequence.

Specifically, time tokens are crucial 
 for positioning events correctly along the entire timeline. 
They benefit from a global attention view
 to compare against all parts of the audio input
 and determine their absolute position.
However, note-on, note-off, and velocity tokens are locally dependent. 
Once a time token has been generated, 
 the subsequent note events are primarily determined 
 by the acoustic features around that specific time. 
These tokens can thus use a more efficient local attention.

We propose a token-type-aware hybrid global-local attention for the encoder-decoder module. 
When the decoder query corresponds to a time token, we compute standard global attention over the entire encoder output. 
When the query corresponds to a note-on, note-off, or velocity token, 
 we apply a local attention mechanism, 
 centered around the time position established by the most recent time token.

Let the output sequence of tokens be denoted by $Y = (y_1, y_2, \dots, y_n)$, where each token $y_i$ has an associated token type $\tau_i \in \{\text{Time}, \text{NoteOn}, \text{NoteOff}, \text{Velocity}\}$.
Each output token $y_i$ is aligned with the input sequence’s temporal position index $t_i$, where $ t_i \in \{1, 2, ..., T\} $ . The core of the mechanism lies in a conditional attention mask:

\begin{equation}
    M_{i,j}^{'} = \begin{cases}
    0 & \text{if  } \tau_i = \text{Time} \\ 
    0 & \text{if  } \tau_i \neq \text{Time},   |j-t_i| \leq w \\
    
    -\infty & \text{otherwise} 
    \end{cases}.
    \label{eq:hybrid-global-local-attn-mask}
\end{equation}

The hybrid global-local attention mechanism 
 leads to significant computational savings
 while retaining the necessary global context for temporal alignment.



\subsection{Hierarchical Pooling}

The sliding window and hybrid global-local attention mechanisms 
 significantly reduce the computational cost. 
However, the cross-attention bottleneck can be further optimized. 
The length of the encoder output sequence, 
 which serves as the key and value for cross-attention, 
 is a critical factor.

We thus introduce a hierarchical pooling scheme 
 between the encoder and decoder. 
Instead of the full-length encoder output $Z$,
 we feed a summary of $Z$ to every decoder layer,
 which is given by applying average pooling 
 with different kernel sizes (temporal resolutions) 
 along the time axis of $Z$.
As shown in Fig.~\ref{fig:efficient-transformer}, 
 the lower layers of the decoder, 
 which handle more foundational predictions, 
 use a heavily pooled (low-resolution) version of the encoder output. 
As we move to higher decoder layers
 that refine the predictions, 
 the pooling size is gradually reduced, 
 providing a higher-resolution context. 
For example, the first two decoder layers 
 might use a pooling size of 4, 
 the next two a pooling size of 2, 
 and the final two layers might use the original, 
 unpooled encoder output.

This strategy offers a trade-off between efficiency and performance. 
It drastically reduces the sequence length 
 for attention computation in the initial decoder layers, 
 where most of the computational cost lies, 
 while providing the final layers
 with high-resolution information to ensure accuracy. 
Our experiment shows this hierarchical approach 
 is more effective than using a single, 
 fixed pooling size across all layers.


\section{Evaluation}

This section reports the experiments
 conducted for evaluating the effectiveness 
 of the proposed sparse attention mechanism
 in terms of performance and efficiency.

\subsection{Dataset}

We used the MAESTRO v3.0.0 dataset \cite{maestro} for all our experiments. 
It is a large-scale dataset containing over 200 hours of virtuosic piano performances, 
 captured on a Yamaha Disklavier. 
The dataset is particularly suitable for this task due to its high-accuracy audio-to-MIDI alignments (with a reported tolerance of ±3ms). 
We used the official train, validation, and test splits provided with the dataset.



\subsection{Configurations}

The input to our model is a mel spectrogram. The audio was first resampled to 16,000 Hz. 
The spectrogram was computed with a hop length of 320 samples (20 ms), Fast Fourier Transform (FFT) window size of 2048, minimum frequency of 20 Hz, maximum frequency of 7600 Hz, and 512 mel frequency bins. 
For training, the audio files were clipped into 10.24-second segments. 
This choice was primarily due to GPU memory constraints 
 for the baseline transformer model.
To ensure fair comparison across all models
 and maintain reproducibility,
 we therefore adopted 10.24-second segments as the standard input length.
The output was the sequence of MIDI-like tokens described in Section 3.2.

All models were implemented in PyTorch\footnote{www.pytorch.org}. We set the encoder layers and decoder layers to 6, hidden embedding size to 512, attention head number to 8, dimension of feed forward layers to 1024, the sliding window size $w$ to 64, and dropout rate to 0.1.
We trained each model for 400,000 steps with a constant learning rate of 1e-4. We utilized the AdamW \cite{adamw-iclr-LoshchilovH19} optimizer with weight decay regularization and a batch size of 32 for training. 
All experiments were conducted on a single NVIDIA RTX 3090 GPU with 24GB of VRAM.


\subsection{Evaluation Metrics}

We evaluated the transcription performance using the standard metrics from the $mir\_eval$ \cite{mir_eval} library . 
We firstly computed the F1-score to assess note accuracy 
 based on correct pitch and an onset time within a 50 ms threshold. 
The second metric adds a condition for the note offset,
 which must be within the greater of 50 ms or 20\%  
 of the ground truth note's duration. 
The third metric requires all previous conditions to be met,
 plus a note velocity that is within 10\% of the ground truth velocity.
 
\subsection{Results}

\begin{table*}[tb]
    \centering
    \caption{Note-Level Transcription Performance F-score of Transformer-Based Seq2Seq Models on the MAESTRO V3.0.0 Test Split.}
    \label{tab:transcription_results}
    \begin{tabular}{@{}lccccccccc@{}}
    \toprule
    \textbf{Model} & \textbf{Enc. Self-Attn.} & \makecell{ \textbf{Enc. Dec. } \\ \textbf{Cross-Attn.}} & \textbf{Dec. Self-Attn.} & \textbf{Pooling} & \textbf{Onset} & \textbf{Onset\&Offset} & \makecell{ \textbf{Onset,Offset} \\ \textbf{\& Velocity} } \\
    \midrule
    Transformer \cite{transformer} & Full & Full & Full & - & 96.01 & 83.46 & 82.18 \\
    \midrule
    Transformer (Reproduced)  & Full & Full & Full & - & \pmb{96.86} & 83.21 & 82.61 \\
    Efficient Transformer V1  & Local & Full & Full & - & 96.72 & \pmb{83.71} & \pmb{83.08} \\
    Efficient Transformer V2  & Local & Full & Local & - & 96.76 & 83.06 & 82.39 \\
    Efficient Transformer V3  & Local & Global+Local & Local & - & 96.57 & 82.82 & 82.18 \\
    Efficient Transformer V4  & Local & Global+Local & Local & 4,4,4,4,4,4 & 96.19 & 81.50 & 80.77 \\
    Efficient Transformer V5  & Local & Global+Local & Local & 4,4,2,2,1,1 & 96.60 & 82.76 & 82.10 \\
    \bottomrule
    \end{tabular}
    \label{tab:experiment-result-baseline}
\end{table*}

\subsubsection{Comparison with Baseline}
We first established a baseline by re-implementing the T5 seq2seq model used in \cite{transformer} with full attention. 
The results are summarized in Table 1.
Our baseline model (reproduced) achieved  scores of 96.86\% (Onset F1 score), 83.21\% (Onset\&Offset F1 score), and 82.61\% (Onset,Offset\&Velocity F1 score). 
These results are closely comparable to the original benchmark scores from \cite{transformer}, 
 which validating our baseline implementation.
 
We then progressively added our proposed efficiency improvements
 and measured the impact on performance. 
As shown in Table \ref{tab:experiment-result-baseline}, the proposed improvements are denoted as efficient Transformer V1 to V5.
Replacing the encoder self-attention with local attention (V1) slightly improved performance
 on the more complex metrics (83.71\% Onset\&Offset F1 score, 83.08\% Onset,Offset\&Velocity F1 score)
 with a negligible impact on the Onset F1 score.
Extending local attention to the decoder (V2) 
 or using a hybrid Global+Local cross-attention (V3)
 did not yield further improvements and resulted in performance closer to the baseline.
Introducing a uniform pooling strategy (V4) significantly degraded performance in all metrics.
However, a hierarchical pooling scheme (V5), which applied less pooling in deeper layers, successfully balanced efficiency and accuracy, recovering performance to a level competitive with the baseline.
As the results indicate, the step-by-step introduction of sparse attention mechanisms leads to a very slight decrease in the F1-scores across all metrics. 
Our final proposed efficient Transformer V5 model maintains performance highly
 comparable to the full-attention baseline.

\subsubsection{Ablation Study}

\begin{figure}[htbp]
\centering
    
    \includegraphics[width=9cm]{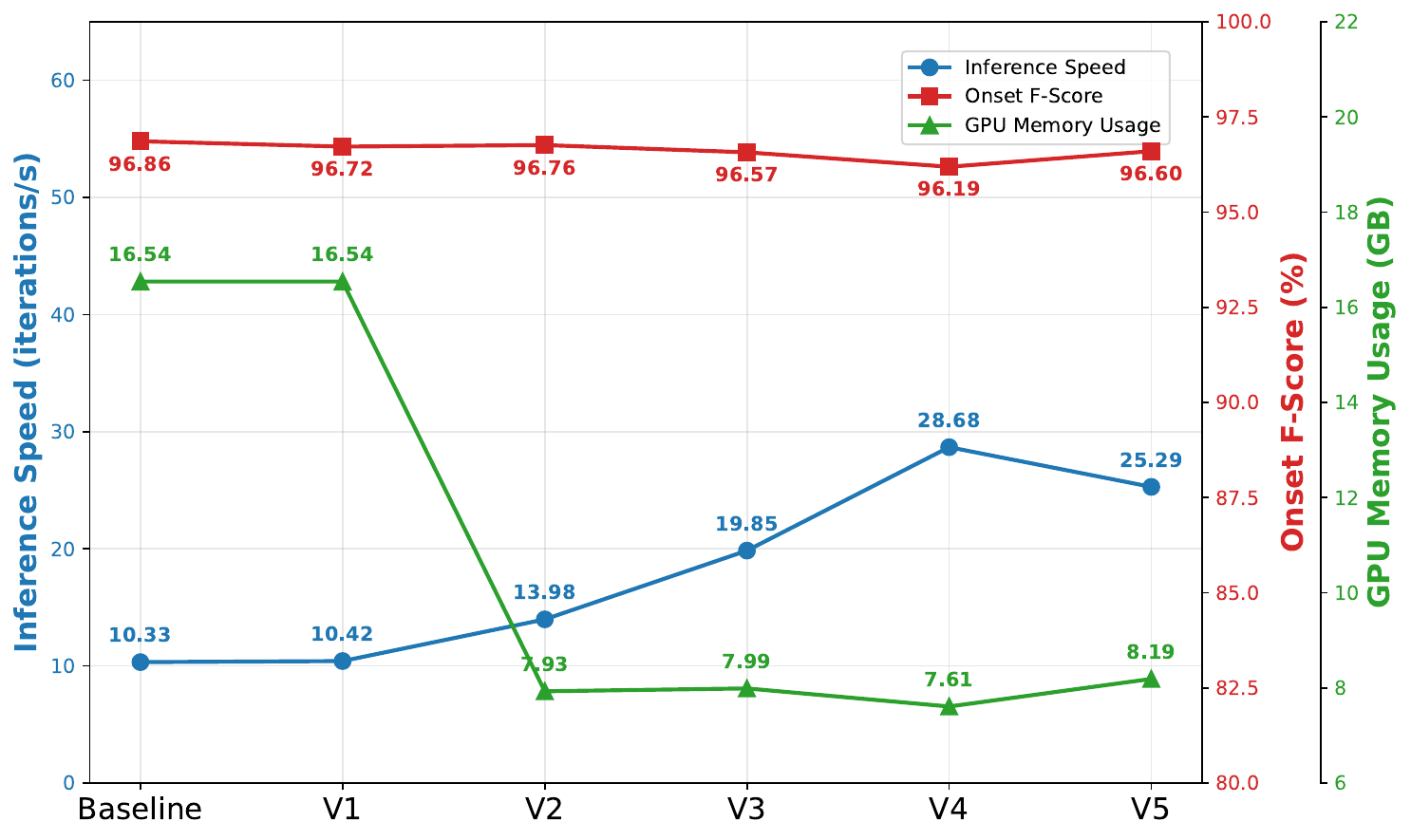}

\caption{Transcription speed and performance for different model configuration.}
\label{fig:inference-speed}
\end{figure}

We evaluated the efficiency of our model in terms of inference speed and maximum GPU memory consumption
 and evaluating their impact on performance. 
We used decoder key-value cache for inference acceleration, set the inference batch size to 256, and maximum output sequence length to 1024.
The result is demonstrated in Fig.~\ref{fig:inference-speed}.
Starting from the baseline model, we first introduced local attention mechanism in the encoder (V1), 
 which has negligible impact. 
Adding local attention mechanism to the decoder (V2) significantly reduces GPU memory usage with minimal accuracy drop. 
Incorporating hybrid global-local attention mechanism for encoder-decoder interaction (V3) further boosted inference speed. 
Applying uniform pooling (V4) led to the highest speed gain but at the cost of a noticeable performance drop. 
Finally, replacing uniform pooling with hierarchical pooling (V5) achieves a better trade-off,
 recovering most of the accuracy while maintaining high efficiency.
The results clearly demonstrate the efficiency of our proposed model.

\begin{table}[tb]
    \centering
    \caption{Model performance (F1 score) with different sliding window sizes.}
    \label{tab:ablation-sliding-window-sizes}
    \begin{tabular}{@{}cccccc@{}}
    \toprule
    \textbf{Sliding Window Size}   & \textbf{Onset} & \textbf{Onset\&Offset} & \makecell{ \textbf{Onset,Offset} \\ \textbf{\& Velocity} } \\
    \midrule
    16  & 95.83 & 80.16 & 79.43 \\
    32  & 96.26 & 82.19 & 81.47 \\
    64  & \pmb{96.60} & 82.76 & 82.10 \\
    128  & 96.51 & \pmb{83.19} & \pmb{82.53} \\
    \bottomrule
    \end{tabular}
    
\end{table}

We also performed an ablation study to determine the optimal sliding window size, with the results summarized in Table \ref{tab:ablation-sliding-window-sizes}.
The data indicates a general trend of improved F1 scores as the window size increases from 16 to 128 across all task configurations.
Although a window size of 128 achieved the highest scores on average, we selected a window size of 64 for our model. A sliding window size of 64 represents the most effective balance between high performance and computational efficiency.

\section{Conclusion}


In this paper, we proposed an efficient Transformer-based model
 for sequence-to-sequence piano transcription. 
Our work is motivated by the key finding that the attention mechanisms
 in standard Transformer models exhibit a high degree of sparsity
 when applied to the piano transcription task. 
We introduced a series of optimizations,
 including sliding window self-attention, 
 a novel hybrid global-local cross-attention, 
 and a hierarchical pooling scheme, 
 to create a more computationally tractable model.

Our experimental results on the MAESTRO dataset confirm our hypothesis.
The proposed sparse attention mechanisms achieve more than 2$\times$ speedup
 in inference and a significant reduction in GPU memory usage.
This efficiency is achieved with only a negligible impact on transcription accuracy.

This work validates the effectiveness of applying sparse attention principles, 
 tailored to the specific characteristics of the music transcription task. 
Future work could explore applying these techniques to other instruments, 
 investigating more dynamic or learned sparse attention patterns,
 and further optimizing the model for real-time applications.

\section*{Acknowledgment}

This work was partially supported by JST FOREST Grant No. JPMJFR2270 and JSPS KAKENHI Grant Nos. 24H00742, 24H00748, 25K22841, and 25H01142.











\bibliographystyle{IEEEtran}
\bibliography{IEEEabrv, mybib}

\end{document}